\documentclass{article}
\usepackage{ijcai23}

\usepackage{times}
\usepackage{soul}
\usepackage{url}
\usepackage[hidelinks]{hyperref}
\usepackage[utf8]{inputenc}
\usepackage[small]{caption}
\usepackage{graphicx}
\usepackage{amsmath}
\usepackage{amsthm}
\usepackage{booktabs}
\usepackage{algorithm}
\usepackage{algorithmic}
\usepackage[switch]{lineno}

\usepackage{float}
\usepackage{listings}
\usepackage{amssymb}
\usepackage{amsfonts}
\usepackage{bm}
\usepackage{multicol}
\usepackage{subfigure}
\usepackage{url}
\usepackage{booktabs} 
\usepackage{color}
\usepackage{arydshln} 
\usepackage{appendix}
\usepackage{bm}
\usepackage{stfloats}

\title{Boosting Value Decomposition via Unit-Wise Attentive State Representation for Cooperative Multi-Agent Reinforcement Learning}

\author{
Qingpeng Zhao$^1$\and
Yuanyang Zhu$^1$\and
Zichuan Liu$^1$\and
Zhi Wang$^1$\and
Chunlin Chen$^1$
\affiliations
$^1$Department of Control Science and Intelligence Engineering,\\ Nanjing University, China\\
\emails
\{qingpeng, yuanyang, zichuanliu\}@smail.nju.edu.cn,
\{zhiwang, clchen\}@nju.edu.cn
\thanks{corresponding to Yuanyang~Zhu and Chunlin~Chen, yuanyang@smail.nju.edu.cn, clchen@nju.edu.cn.}
}
\begin{document}

\maketitle

\begin{abstract}
   In cooperative multi-agent reinforcement learning (MARL), the environmental stochasticity and uncertainties will increase exponentially when the number of agents increases, which puts hard pressure on how to come up with a compact latent representation from partial observation for boosting value decomposition.
   To tackle these issues, we propose a simple yet powerful method that alleviates partial observability and efficiently promotes coordination by introducing the UNit-wise attentive State Representation (UNSR).
   In UNSR, each agent learns a compact and disentangled unit-wise state representation outputted from transformer blocks, and produces its local action-value function. 
   The proposed UNSR is used to boost the value decomposition with a multi-head attention mechanism for producing efficient credit assignment in the mixing network, providing an efficient reasoning path between the individual value function and joint value function.
   Experimental results demonstrate that our method achieves superior performance and data efficiency compared to solid baselines on the StarCraft II micromanagement challenge. 
   Additional ablation experiments also help identify the key factors contributing to the performance of UNSR.
    
\end{abstract}

\section{Introduction}

Cooperative multi-agent reinforcement learning (MARL) has great promise for addressing complex real-world problems, such as in autonomous driving~\cite{car}, sensor networks~\cite{zhang2011coordinated} and robotics control~\cite{huttenrauch2017guided}.
Centralized training with decentralized execution (CTDE) becomes a popular paradigm in cooperative MARL to promote collaboration among agents, where the independent agents can access global state information that is unavailable during their policy inference.
However, it also brings an intractable challenge which is how to efficiently represent and use the action-value function that all agents learn.
Especially, the environmental stochasticity and uncertainties will increase exponentially when the number of agents increases.

Typically, credit assignment in value decomposition~\cite{vdn,qmix} is a crucial aspect of MARL as it allows for determining the relative contributions of different agents to the overall success.
Since the more accurately decomposed individual Q-values could update the policies more accurately, it is essential for guiding the learning of decentralized policies, as it allows the agents to focus on improving their own performance while still working towards achieving overall success.
One common approach to addressing credit assignments in MARL is to embed it as a module with an attentively designed central mixer, where the mixer generally conditions on global state and 
individual value functions of each agent~\cite{vdn,qmix,qatten,qplex,rashid2020wqmix}.
However, the above works generally leverage all available information for estimating the joint value function, and ignore the multi-agent characters that the observation consists of a number of object entities, which can be quite inefﬁcient.
Unlike previous works that restrict the representation relation of the individual Q-values and the global one, QPD~\cite{qpd} compute the credits using action-observation history in the mixer.
From the causal inference perspective, DVD~\cite{dvd} introduces the trajectory graph depending only on the local trajectories to represent the local contributions.
However, none of the above works explicitly leverage the fact that individual local information may not only have different influences on its individual value function but also affect the credit assignment.
Here, we use unit-wise attentive state representation to model the individual value function of each agent while the state representation is used to establish the relations from the individuals to the global value function for boosting value decomposition.

For the problems of non-stationarity and partial observability, an appealing paradigm is Centralized Training and Decentralized Execution (CTDE).
Many CTDE methods focus on using a specific network structure to relieve the partial observability, most previous works use memory architecture, such as recurrent neural network (RNN), long-short term memory (LSTM)~\cite{hochreiter1997long} and GRU~\cite{chung2014empirical}, to help the agent effectively capture long-term dependencies conditioning on their entire action-observation history.
In this case, gated Recurrent Neural Networks (RNNs), has been served as the most popular mechanism to store additional information from historical observations over longer timescales in the MARL community~\cite{vdn,qmix,qplex}.
Recently the transformer architecture has shown superior performance over RNN in capturing multi-step temporal dependencies to deal with the issue of partial observability~\cite{parisotto2020stabilizing,yang2022transformer,mat}.
However, the transformer-based methods remain limited since the memory computation surges when the historical observation becomes complex.
Moreover, they also ignore the processing of individual functions on different entities and receive limited representation during the learning process.
In this paper, we leverage the light encoder module of the transformer to produce a unit-wise state representation for each agent that contains meaningful and clear relationships between entity interactions.

End-to-end MARL learns compact and informative representations implicitly as part of the learning process, which is receiving signiﬁcant attention.
One challenge of the representation problems is how we can give full play to agents’ specialization given its action-observation history for improving learning efﬁciency.
ROMA~\cite{roma} constructs a stochastic role embedding space to enable adaptive shared learning among agents.
RODE~\cite{rode} learns a role selector based on action effects and integrates it into the role policies to improve learning efﬁciency.
LINDA~\cite{linda} moves a step towards leveraging local information to build awareness for each teammate by learning to decompose local information in their local networks. 
ASN~\cite{asn} utilizes action semantics between agents that different actions have different influences on other agents, resulting in more efficient coordination.
While they use the semantics of local information to model the individual value function efficiently, they do not explicitly utilize the local semantics to directly promote factorizing the joint value function.

To address the problems above, we propose a novel method called UNit-wise attentive State Representation (UNSR) that leverages attentive state representation for each agent to improve learning efﬁciency and draws it into value decomposition to boost credit assignment.
Instead of using gate RNNs to tackle partial observability for each agent, we attempt to leverage the transformer encoder as a memory process structure to generate a unit-wise state representation for capturing the internal relationships of historical observation.
Specifically, each agent model its value function via feeding their unit-wise state into a multi-layer perceptron (MLP), where the unit-wise state conditions on its own action-observation history $\tau$.
In this case, UNSR can effectively avoid the negative influence of irrelevant input information, and thus provide a more accurate estimation of its value function.
Then, we concatenate the unit-wise state representation of each agent together with individual value functions and global state to estimate the joint value function via a multi-head attention mechanism.
Here, we believe that sufficient state representation information can help the mixing network deduce how important the role each agent plays in cooperative tasks is.
It means that conditioned on sufficient information, the mixing network for the credit assignment should get aligned and consistent in the unit-wise state representation space.

We benchmark UNSR on a range of StarCraft Multi-Agent Challenge (SMAC) scenarios, and observe considerable performance improvements over the state-of-the-art baselines, especially on the hard and super-hard scenarios.
We also conduct further component studies to demonstrate the effectiveness of unit-wise state representation for individual function and value decomposition, which confirms that each component is a key part of UNSR.

\section{Preliminaries}
\textbf{Multi-agent Markov Decision Process.}
A fully cooperative multi-agent task is an extension of a decentralized partially observable Markov decision process (Dec-POMDP)~\cite{oliehoek2016concise}, which is modeled by a tuple $G = \langle\mathcal{N}, \mathcal{S}, \mathcal{A}, \mathcal{P}, \Omega, O, r, \gamma\rangle$
, where $\mathcal{N}$ represents a finite of agents with $\mathcal{N} \equiv\{1,2, \ldots, n\}$, and  $s \in \mathcal{S}$ is a global states of the environment. 
At each time step $t$, each agent $i\in \mathcal{N}$ selects an action  $a_{i} \in \mathcal{A}$ to formulate a joint action, i.e., $\boldsymbol{a} \equiv\left[a_{i}\right]_{i=1}^{n} \in \mathcal{A}^n$. 
The result of executing this joint action is that all agents receive a shared reward according to the reward function $r(s, \boldsymbol{a})$, and then observe a new state transition on the environment depending on the transition probability function $s^{\prime} \sim \mathcal{P}(\cdot \mid s, \boldsymbol{a})$.
Moreover, considering in a partial observation setting, each agent $i$ only has access to an individual observation $o_{i} \in \Omega$, which is derived by the observation probability function $O\left(o_{i} \mid s, a_{i}\right)$. 
Each agent $i$ has action-observation history $\tau_{i} \in \mathcal{T} \equiv \Omega \times \mathcal{A}$ that conditions a stochastic policy $\pi_{i}$, and the joint action-observation history is defined as $\boldsymbol{\tau} \in \mathcal{T} \equiv \mathcal{T}^{n}$.
The goal is to find a joint policy $\boldsymbol{\pi}=\left\langle\pi_{1}, \ldots, \pi_{n}\right\rangle$, inducing the following joint action value as  $Q^{\boldsymbol{\pi}}(s, \boldsymbol{a})=r(s, \boldsymbol{a})+\gamma \mathbb{E}_{s^{\prime}}\left[R_t|s, \boldsymbol{a}\right]$, where $\gamma$ is a discount factor and $R_t=\sum_{t=0}^{\infty} \gamma^t r_t$ is the cumulative return.

\textbf{Centralized Training with Decentralized Execution.} CTDE is a common architecture used in the MARL, where each agent learns a policy only on its own action observations and the centralized value function provides a centralized gradient to update the individual function.
One of the promising ways to exploit the CTDE framework is value decomposition, which allows agents to learn their individual utility functions by optimizing the joint action-value function for credit assignment.
To ensure consistency~\cite{qtran} for multi-agent value decomposition methods, it should satisfy the IGM (Individual-Global-Max) principle:
\begin{equation}
\label{dsdsadad}
\operatorname{argmax}_{\boldsymbol{a}} Q_{tot}(\boldsymbol{\tau}, \boldsymbol{a})
=\left(\begin{array}{c}
\operatorname{argmax}_{a_{1}} Q_{1}\left(\tau_{1}, a_{1}\right) \\
\vdots \\
\operatorname{argmax}_{a_{n}} Q_{n}\left(\tau_{n}, a_{n}\right)
\end{array}\right),
\end{equation}
where $\boldsymbol{a}$ is the joint action-observation history.

\textbf{Attention Mechanism in Transformers.}
Harnessing the transformers to handle partial observability is at the core of the attention mechanism, which has been investigated~\cite{maac,qatten} in MARL.
An attention function adopts three matrices representing a set of queries $Q$, keys $K$, and values $V$.
Mathematically,  the attention  is computed as a weighted sum of the values as
\begin{equation}
\operatorname{Attention}(Q,K,V)=\operatorname{softmax}(\frac{Q K^\top}{\sqrt{d_{k}}})V,
\end{equation}
where  $\frac{1}{\sqrt{d_k}}$ denotes a scaling factor.
In practice, a multi-head structure is often used to enable the model to focus on different subspaces of representations. 
Thus, the multi-head attention in transformers utilizes the set of $m$ projections as
\begin{equation}
\begin{split}
\operatorname{MHA}=\operatorname{Concat}\left(\operatorname{head}^{1}, \ldots, \operatorname{head}^{m}\right) W_{\operatorname{MHA}},
\end{split}
\end{equation}
where $ \{\operatorname{head}^{1}, \ldots, \operatorname{head}^{m}\}$ represent a set of parallel attention mechanisms and $W_{\operatorname{MHA}}$ is the learnable weight for concatenating all vectors from each attention head.

\begin{figure*}[tb]
   \centering
    \includegraphics[width=2\columnwidth]{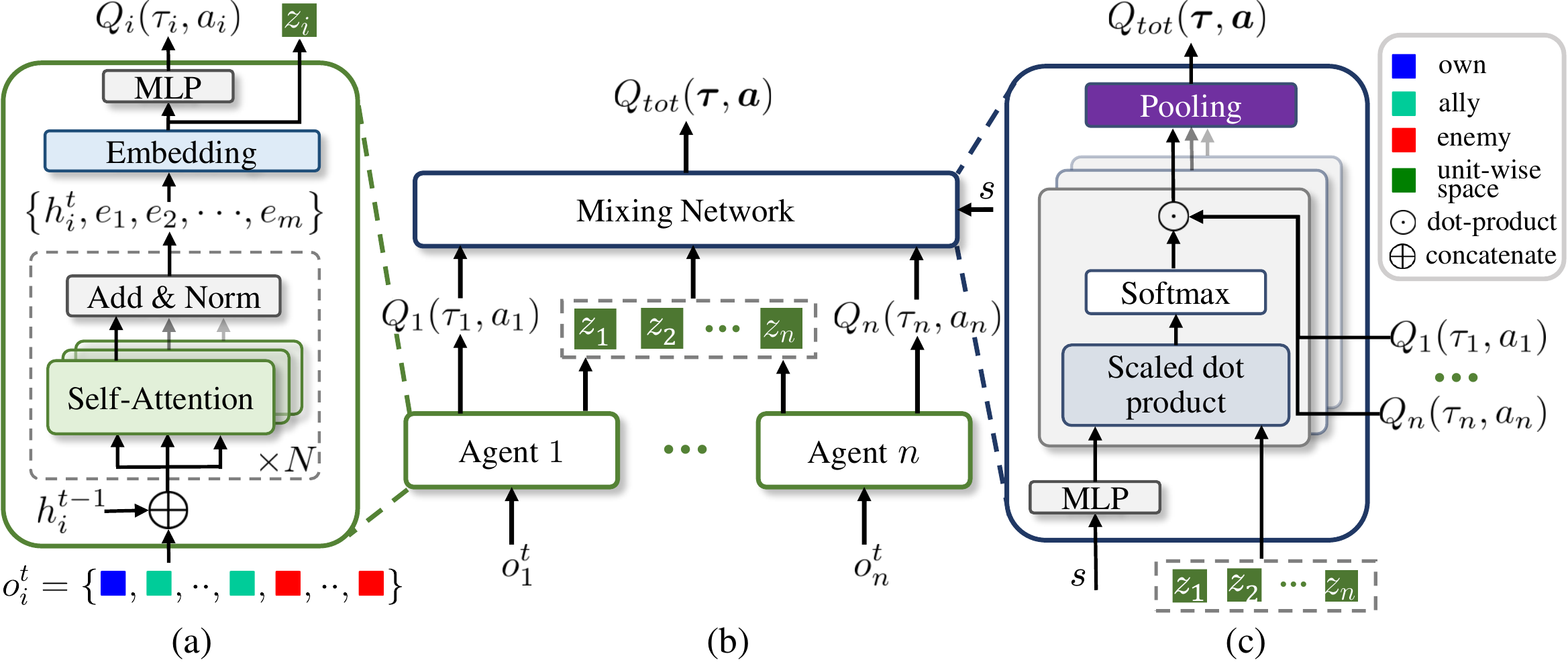}
    \caption{The framework of our method UNSR. 
    First, each agent models a value function condition on its local observation-action history.
    Then, we construct a unit-wise state representation $\mathcal{Z}_i$ via embedding the outputs of their computational blocks.
    The right is the mixing network of UNSR, where unit-wise attentive state representation $\mathcal{Z}$ together with global state $s$ and local value functions $\left\{Q_1, \ldots, Q_n\right\}$ are used to estimate the joint value function.
    ``$\times N$" means that there are L layers for the transformer block.
    }
    \label{framework}
\end{figure*}

\section{Method}

\subsection{Motivation}
As a popular class of CTDE algorithms, value decomposition methods show their strength in expressing the joint value function conditioned on global trajectory using individual value functions conditioned on the local observation-action history.
While CTDE can reduce the number of learnable parameters and growing joint-action space complexity, it still faces a critical challenge of credit assignment, especially when the number of agents increases, i.e., how to tease out a compact representation of the underlying task-relevant information from which to learn instead given high-dimensional inputs.
Moreover, an inefficient representation could fail to deduce the contributions of individual agents given only global rewards.
Previous works have made a pave to improve representation expressiveness of high-level state information or value function by either incorporating effective and expressive network design~\cite{mat,updet,qatten} or designing specialized network structures to learn semantic information~\cite{dvd,asn}.
However, previous works usually use all available information for estimating the value function of each agent, which could be quite inefficient for facilitating multi-agent coordination.
And, none of them explicitly considers extracting local information with high-level representation, which is a critical factor we can leverage to efficiently estimate the value function and facilitate coordination, especially in challenging multi-agent tasks.

To this end, we propose a simple yet powerful value decomposition method, called unit-wise attentive state representation (UNSR), which learns the unit-wise attentive state representation to alleviate partial observability while using it to efficiently promote coordination.
Over the course of training, each agent teases out the relevant information into a compact and disentangled unit-wise state representation outputted from transformer blocks, which produces its local action-value function conditioning on its local observation and hidden state.
In the mixing network, the multi-head attention mechanism is used to model the individual impact on the whole system at the agent level, where unit-wise state representation $\mathcal{Z}$ together with global state $s$ and local value functions $\left\{Q_1, \ldots, Q_n\right\}$ are gathered to estimate the joint value function.
In this way, UNSR can effectively avoid the negative influence of irrelevant information with unit-wise attentive state representation, thus outputting stable and consistent action-value functions and efficiently promoting credit assignment for each agent.
Figure~\ref{framework} provides an overview of our UNSR framework.

\subsection{The Practical Implementation of UNSR}
Based on the previous analysis, we will give the implementation details of UNSR.
First, each agent learns a unit-wise state in terms of a compact representation through a transformer encoder module composed of a stack of $N$ identical layers, which will be further passed to the individual value function- and mixing- network, respectively.
Then, we model the value function with the obtained unit-wise attentive state representation for each agent.
Afterward, the unit-wise and global states are fed to the mixer to deduce the contributions of each agent with the multi-head attention mechanism.
Finally, the joint value functions are predicted depending on the credits together with the individual value function of each agent.

\textbf{Unit-wise Attentive State Representation.} Here, we give the detailed design of the unit-wise state with compact and disentangled representation by using the transformer encoder module.
Each agent maintains its unit-wise state representation, produced conditioning on its local observation and hidden state information at each timestep.

Considering the semantic difference of the various observation space, we divide entities in the agent observation space of the agent $i$ into three spatial entity sets: $o_{\text {self }}^i$, $o_{\text {ally}}^i$ and $o_{\text {ent}}^i$.
The self entity $o_{\text {self }}^i$, entities of other allied agents $o_{\text {ally}}^i$ and remaining entities $o_{\text {ent}}^i$ represent the agent $i$'s individual features, agent $i$'s observation to each other agents and the others information (such as enemies or environmental objects).
Before feeding these entities into the transformer block, to handle the various observation space and enable the hidden state space, these entities are input together with the hidden state into the transformer block after being embedded into $\left\{e_{i,1}^{in}, e_{i,2}^{in},\cdots, e_{i,n}^{in}\right\}$ via an embedding layer.

As shown in Figure~\ref{framework}-(a), the unit-wise state representation module contains $N$ transformer blocks which are used to generate the outputting entity embeddings $f_i^{t}= \left\{h_i^t, e_1^{out}, e_2^{out}, \cdots, e_m^{out}\right\}$ consisting of a history state information $h_i^t$ and $m$ entities $e$ given the above observation semantic embedding and hidden historical states as input, and an embedding module which generates the unit-wise attentive state representation $\mathcal{Z}_i^{t}$ as output.
In this case, the generated history state information $h_i^t$ together with the new observation semantic embedding will be fed into transformer blocks at the next timestep.
Each such transformer block consists of a self-attention mechanism and an MLP, as well as residual connections to prevent gradient vanishing and network degradation with the increase of depth.
To better tradeoff between performance and memory consumption, we generally utilize $2\sim4$ stacked identical layers as an encoder in implementation.
Instead of using RNN, we utilize transformers as an encoder technique not only aiming to capture multi-step temporal dependencies to deal with the partial observability by hidden historical states, but also to capture the interrelationship of input elements with a compact latent state representation.
Then, each agent's unit-wise state representation $\mathcal{Z}_i^{t}$ is used to model its value function and be fed into the mixer for approximating its coefﬁcients and estimating its contributions to the team, respectively. 

In practice, we first embed the individual input observation $o^t_i$ of agent $i$ as $\left\{e_{i,1}^{in}, e_{i,2}^{in},\cdots, e_{i,n}^{in}\right\}$, then concatenate it with the hidden state $h^{t-1}_i$ producing matrix $f_{i,1}^t=\{h^{t-1}_i, \textbf{e}^t_i\}$, which are packed together into three matrices, $Q_i=W_qf_{i,l}^t$, $K_i=W_kf_{i,l}^t$, and $V_i=W_vf_{i,l}^t$ representing a set of queries, keys, and values respectively, where $l\in \{1,\cdots, N\}$ is the number of transformer layers. 
Thus, we formulate the matrix of outputting embeddings as
\begin{equation}\label{ep2}
\begin{split}
\operatorname{Attention}(Q_{i},K_{i},V_{i})=\operatorname{softmax}(\frac{{Q_{i}}{K_{i}^{\top}}}{\sqrt{d_{k}}})V_{i},\\
f_{i,l}^t=\operatorname{LayerNorm}(\operatorname{Attention}(Q_i,K_i,V_i)+f_{i,l-1}^t),
\end{split} 
\end{equation}
where $\frac{1}{\sqrt{d_k}}$ denotes a scaling factor from the dimension of $K_i$ and $\operatorname{LayerNorm}$ is an operation of layer normalization for accelerating training convergence. 
After the transformer block updates all entity embeddings, we denote the outputting entities embeddings $f_{i}^t$ of the last transformer layer as $f_{i}^t = f_{i,N}^t = \left\{h_i^t, e_1, e_2, \cdots, e_m\right\}$. 
Then, the outputting entities embeddings $f_i^{t}$ are used to produce the unit-wise state $\mathcal{Z}_i^t$ as timestep $t$ through an embedding layer $\operatorname{Emb}$ as 
	\begin{equation}
	\mathcal{Z}_i^t = 
 \operatorname{Emb}(f_i^{t}),
	\end{equation}
where $Emb$ is a linear layer with a Relu activation function.

\textbf{Individual Action-Value Function.}
For each agent, the unit-wise state representation $\mathcal{Z}_i^t$ is fed into an MLP as an action-value network to produce action-values $Q(\tau_{i},a_i)$, where the unit-wise state $\mathcal{Z}_i^t$ takes its own action-observation history $\tau$ as input.
This action-value network is used for each agent to determine its own action by calculating the action value given the action-observation history $\tau_i$.

\textbf{Joint Action-Value Function.}
Under the CTDE setting in MARL, each agent must learn to coordinate with each other only given its private observations.
Non-stationarity and partial observability will increase exponentially when the number of agents increases.
It brings a great challenge to credit assignment that aims to deduce the contributions of individual agents from the overall success.
To relieve the above issues, we also investigate value decomposition from the perspective of causal inference.
Then, we ask the following question: can we design a space representation that not only could capture the relationship of input elements to model its local value function but also produce efficient credit assignment for selecting the most appropriate coordination patterns?

Inspired by~\cite{dvd,asn}, we introduce the unit-wise state representation of each agent into the mixing network to estimate the joint value function together with all individual value functions and global state.
Here, we believe that an efficient unit-wise state representation can capture high-level local information semantics which is a critical factor to facilitate the estimation of its value function.
Besides, the high-level space representation can be beneficial for the mixing network to produce efficient credit assignments with the most relevant information.
From the perspective of causal inference, we relieve the spurious correlation between the global state $s$ and joint value function $Q_{tot}$ by incorporating the high-level space representation of each agent into the global state $s$ to learn the credit assignment.

To deduce the contribution of each agent to the team while alleviating the computational burden, we utilize a multi-head attention mechanism to approximate the coefﬁcients and produce the relations from the individuals to the team by selectively paying attention to the unit-wise state from the global state.
Specifically, we compute the similarity weight between the global state $s$ and the unit-wise state representation $\mathcal{Z}$ into a softmax function, which they use as key and query, respectively.
We then gather Q-values for the specific action position from each agent and concatenate them as attention values of the model, and maintain consistency in each head's input.
Mathematically, the attentional representation of one of its heads $Q_h$ is
\begin{equation}
Q_h=\operatorname{softmax}\left (\frac{s^{\top}W_s^{\top}{W_z\mathcal{Z}}}{\sqrt{d_{s}}}\right ) \left [ Q_{1}, \cdots, Q_{n} \right ]^{\top},
\end{equation} 
where $W_s$ and $W_z$ are the weights to embed the corresponding vectors for the global state $s$ and unit-wise state representation $\mathcal{Z}$, $\frac{1}{\sqrt{d_s}}$ behaves as a scaling factor from the dimension of $s$, $Q_h$ represents the generated subspace $Q_{tot}$. 
To focus on information from different representation subspaces at different positions, we apply a number $m$ of attention heads to obtain multiple generative vectors in parallel.
Intuitively, multiple attention heads allow for attending to parts of the sequence differently.
The independent attentional outputs by the heads are then concatenated and linearly pooled into the intended dimensions, i.e.,
\begin{equation}
Q_{tot} =\operatorname{Concat}\left[{Q}^1_{h}, \cdots, {Q}^m_{h}\right] W_{m},
\end{equation}
where $W_{m}$, generated through a hyper network, denotes the weight of the multi-head, which takes the state $s$ as input for the linear pooling layer.
To ensure the monotonicity constraint of Eq.(\ref{dsdsadad}), the weight $W_{m}$ is generated by an absolute function to keep it non-negative.

\begin{figure*}[ht]
   \centering
    \includegraphics[width=2\columnwidth]{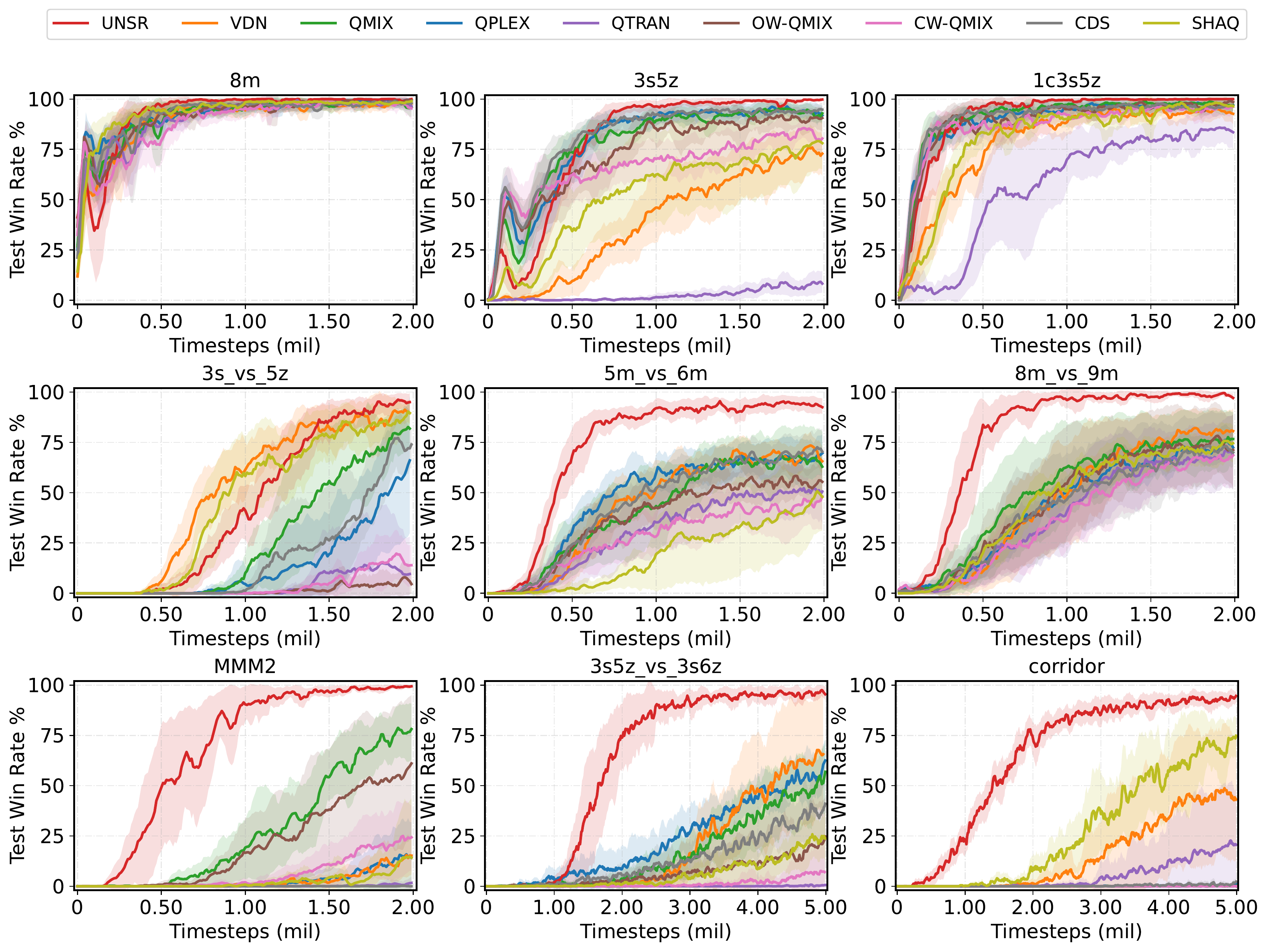}
    \caption{Performance comparison with baselines on easy (first line), hard (second line), and super-hard (last line) scenarios.}
    \label{deepimage}
\end{figure*}

\textbf{Optimization.}
Under the training execution framework of CTDE, UNSR learns by sampling a multitude of transitions from a replay buffer, and we use the standard squared temporal-difference~(TD) error to optimize our whole framework as
\begin{equation}
\mathcal{L}(\theta)=\mathbb{E}_{(\boldsymbol{\tau}, \boldsymbol{a}, r, \boldsymbol{\tau}^{\prime})\in b}\left[\left(y-Q_{tot}(\boldsymbol{\tau}, \boldsymbol{a}; \theta)\right)^{2}\right],
\end{equation}
where $b$ represents a number of transitions by sampling from the replay buffer. 
The TD target value $y=r+\gamma \max _{\boldsymbol{a}^{\prime}} Q_{tot}\left(\boldsymbol{\tau}^{\prime}, \boldsymbol{a}^{\prime}; \theta^{-}\right)$, where $\theta^{-}$ are the parameters of a target network which are periodically copied from $\theta$ and kept constant for a number of iterations.

\section{Experiments}
In this section, we evaluate the performance of our approach compared with three representative value decomposition baselines, VDN~\cite{vdn}, QMIX~\cite{qmix} and QTRAN~\cite{qtran}, and four state-of-the-art MARL algorithms, including WQMIX~\cite{rashid2020wqmix}(both CW-QMIX and OW-QMIX), QPLEX~\cite{qplex}, CDS~\cite{li2021cds} and SHAQ~\cite{wang2021shaq}.
To ensure fair evaluation, we conducted all experiments with five random seeds and the results are presented as means with a 75\% confidence interval, and all algorithm base settings remain consistent.
Note that all algorithms are based on PyMARL framework implementation\footnote{The source code is from~\url{https://github.com/oxwhirl/pymarl}}, and the detailed network structure and hyperparameter settings can be found in Appendix~\ref{appendix2}. 
The experimental results show that UNSR achieves superior performance and learning efficiency compared to all baselines.
Further, we tested in detail the performance of each module in the ablation experiment, aiming to illustrate the advantages of unit-wise state representation.

\subsection{Environmental Settings}
To evaluate the effectiveness of the proposed approach, we carry out experiments with different scenarios on the StarCraft II micro-management challenge (SMAC)~\cite{smac} environment.
SMAC is a popularly used combat benchmark based on real-time strategy, where agents are required to cooperate with each other in micro-management tasks to fight opposing armies controlled by manual coding of built-in AI with unknown strategies.
At each time step, agents perform actions to move or attack any enemy and receive a global bonus regarding the total damage gained.
We tested our method on 12 different maps, including easy, hard, and super-hard scenarios, and the detailed descriptions of this environment can be found in  Appendix~\ref{appendix1}.
Note that our experiments are in the SC2.4.10 version and are not necessarily comparable across versions.

\subsection{Overall Performance Comparison}
As shown in Figure~\ref{deepimage}, UNSR has huge superiority over multiple methods on various scenarios, and achieves close to 100\% win rate for most scenarios with limited training time, especially on super hard scenarios, which validates the effectiveness of our UNSR.
Specifically, UNSR and most algorithms are able to accomplish the task of finding the optimal policy on the easy scenarios (first line), whereas QTRAN performs poorly as its strict convergence conditions may hinder the direction of training optimization.
On the 3s\_vs\_5z map, the key factor for the agents to win is the trade-off between moving and firing to kite the enemies.
UNSR ultimately outperforms SHAQ and VDN by a small margin and is well ahead of the other algorithms.
We suspect that a more sophisticated credit assignment could learn this deeper strategy in team combat.
For hard tasks 5m\_vs\_6m and 8m\_vs\_9m,  UNSR masters the winning strategies through concentrated fire attacks and fewer blood retreats, and outperforms all baselines by a large margin, showing its advantage of unit-wise state representation with environmental stochasticity.
Especially on the super hard scenarios (third line), UNSR improves the average win rate of the state-of-the-art algorithms by almost 40\% and attains the optimal policy with near 100\%.
For example, in the heterogeneous MMM2 map, some existing algorithms (e.g., VDN, QPLEX, and CDS) near-failure work due to unsuccessful value decomposition.
And compared with QMIX and OW-QMIX,  our method converges quickly after $1$ mil training steps and maintains a win rate of more than 90\% with homogeneity for learning sophisticated cooperation.
Also, on challenging 3s5z\_vs\_3s6z and corridor maps, UNSR behaves perfectly, but other algorithms may need help to explore.
It indicates that in such cases, UNSR can use transformer encoders to effectively avoid the negative impact of irrelevant information and thus make a more precise estimate of its policy.
Besides, the compact and disentangled unit-wise state representation produces more accurately decomposed individual Q-values, which also could update the policies more accurately.
In summary, UNSR achieves superior performance and learning efficiency in searching for valuable strategies with consistent updates. 
The full results on all hard and super hard scenarios are shown in Appendix~\ref{appendix3}.

\begin{figure*}[ht]
   \centering
    \includegraphics[width=2\columnwidth]{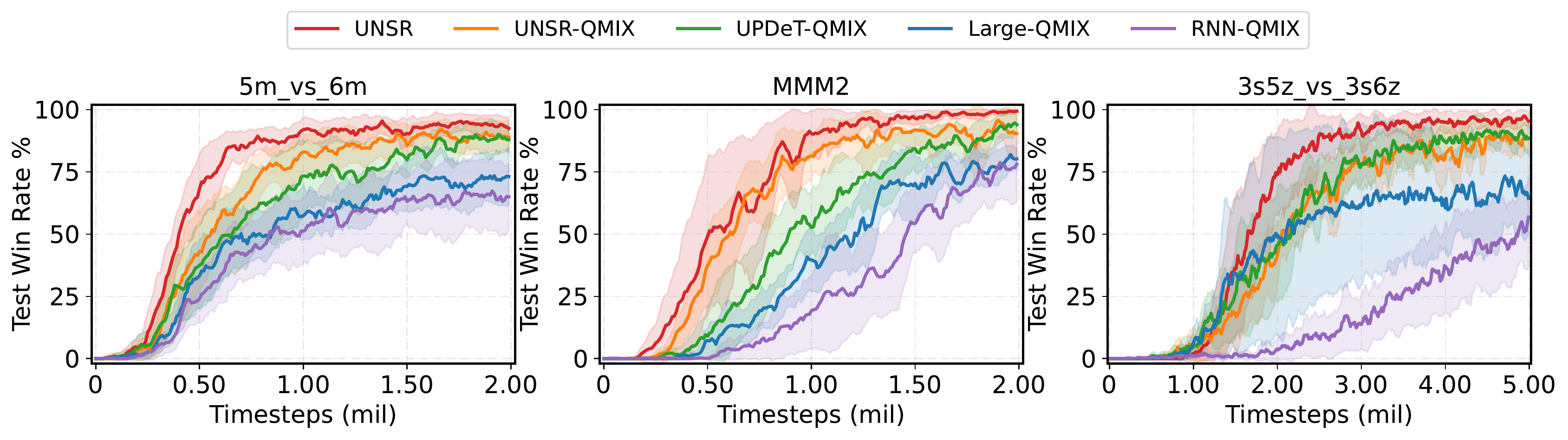}
    \caption{The ablation studies results of UNSR, UNSR with QMIX, UPDeT with QMIX, large hidden dimension RNN network with QMIX and RNN with QMIX on three hard and super hard scenarios.}
    \label{ablation1}
    \vskip -0.1in
\end{figure*}
\begin{figure*}[ht]
   \centering
    \includegraphics[width=2\columnwidth]{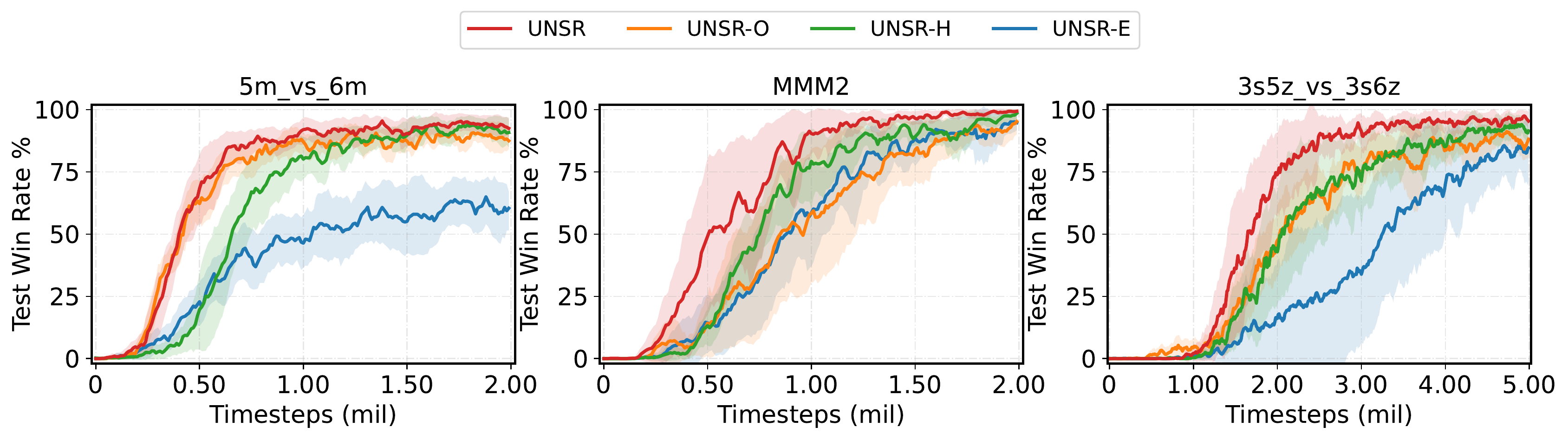}
    \caption{The ablation studies results of UNSR, UNSR-O, UNSR-H, and UNSR-E on three hard and super hard scenarios.}
    \label{ablation2}
    \vskip -0.1in
\end{figure*}

\subsection{Ablation Studies}
To understand the superior performance of UNSR, we carry out ablation studies to test the contribution of its three main components to answer the following questions: (1) Does using wise-unit attentive state representation for individual value function make sense? (2) Does the wise-unit attentive state representation that we introduce into the mixing network the reason our method works?

First, we study the effectiveness of the unit-wise attentive state representation.
We denote the combination of the UNSR and UPDeT individual Q-value function with QMIX joint Q-network as UNSR-QMIX  and UPDeT-QMIX, and the use of RNN networks with hidden dimensions of $64$ and $256$ for the QMIX individual Q-value function as RNN-QMIX and Large-QMIX, respectively.
As shown in Figure~\ref{ablation1}, UNSR-QMIX obtains gratifying results superior to UPDeT-QMIX, Large-QMIX, and RNN-QMIX, indicating that the UNSR module successfully captures the relevant-task input local information with the unit-wise state representation from a larger amount of mixed observation information and plays a signiﬁcant role in enhancing learning performance.
The results show that UNSR-QMIX outperforms UPDeT-QMIX in all tasks, the performance improvement achieved by UNSR should be related to the use of the hidden state information, which could capture long-term dependencies over longtime steps.
Moreover, it is notable that the improvement of incorporating UPDeT into QMIX relies on the representation capability of the transformer than RNN-QMIX and Large-QMIX, which indicates that action-group strategy can improve the performance with policy decoupling strategy and a paired observation entity.
The result also proves that the transformer-based model can outperform the conventional RNN-based model.
Large-QMIX has a slight performance advantage over RNN-QMIX with such a larger network size, which demonstrates that larger networks will not deﬁnitely bring performance improvement, and the superiority of UNSR does not come from the larger networks.
We accordingly conclude that the unit-wise attentive state representation owns the greater representation ability to keep track of the feature inﬂuence of each agent of each kind in the individual value function prediction process, and promote coordination among agents.

Further, to verify whether incorporating the unit-wise attentive representation into the value decomposition network is rational, we replaced unit-wise state representation $\mathcal{Z}_i$ with observation $o_i$, historical hidden state $h_i^{t}$ and $m$ entities for each agent $i$, denoted them as UNSR-O, UNSR-H, and UNSR-E, respectively, where UNSR-E build the unit-wise state space $\mathcal{Z}_i$ only condition on local observation without hidden state information.
As we can see from Figure~\ref{ablation2}, UNSR has a significantly better performance than UNSR-O, UNSR-H, and UNSR-E on all scenarios.
UNSR-O and UNSR-H achieve similar performance, which conﬁrms that the hidden state representation could capture the important information of the entire observation space.
UNSR-E performs worse than UNSR-H, suggesting that the memory structure can promote the long-horizon reasoning ability in partial observation tasks.
When replacing the $\mathcal{Z}_i$ with observation $o_i$ or hidden state $h_i^{t}$ leads to a degradation in learning performance across all tasks, it implies that the attentive state representation can help the mixing network deduce the contribution of each agent and obtain better performance.
This demonstrates that hidden state, agent-speciﬁc features, and global information are all important in forming an effective global state.
Another factor can be that the unit-wise state representation helps the mixer build the relationship between individual value functions and joint value function conditioning on the global and compact high-level state representation.
These observations support the claim that introducing individual information can improve value decomposition efﬁciency and help agents select the most appropriate coordination patterns.

\section{Conclusion}
In this work, we propose UNSR to facilitate more efﬁcient value decomposition by incorporating the unit-wise attentive state representation into the mixer.
Technically, we build the unit-wise attentive state representation with the encoder module of the transformer, which is used to predict the individual value function and deduce the contributions of each agent in value decomposition, respectively.
To the best of our knowledge, UNSR is the ﬁrst to attempt to explicitly build a bridge between individual value function and joint value function in terms of individual state representation at each timestep.
Experiments on the challenging SMAC benchmark show that our method obtains the best performance on almost all maps, and the individual value module of UNSR could boost existing MARL algorithms with impressive learning acceleration and performance improvement when integrated with them.
This simple yet effective UNSR method further motivates us to explore high-quality representation in future work.
Another challenging direction is the theoretical study of the ambiguous credit assignment phenomenon from the perspective of the causal graph.

\bibliographystyle{named}
\bibliography{ijcai23}

\newpage
\onecolumn
\appendix
\setlength{\baselineskip}{2em}

\noindent \vspace{0.5em} \rule{\textwidth}{0.15em}
\centerline{\LARGE \textbf{Supplementary Materials}} \\
\centerline{\LARGE \textbf{Boosting Value Decomposition via Unit-Wise Attentive State Representation }} \\
\centerline{\LARGE \textbf{for Cooperative Multi-Agent Reinforcement Learning}}
\vspace{0.5em} \rule{\textwidth}{0.15em}

\section{Experimental Environment Settings}
\label{appendix1}
In our experiments, we utilized the StarCraft II micromanagement challenge (SMAC), a multi-agent reinforcement learning platform that simulates the real-time strategy game StarCraft, to verify and test the proposed method. 
The SMAC environment is comprised of two distinct levels: macro-operation and micro-operation. 
The macro-operation level involves high-level decision-making and requires a comprehensive approach that integrates resource management, economy, and military strategy in order to win the game. 
On the other hand, the micro-operation level focuses on the local control of agents, where the agents are only able to receive information from a limited perspective and do not have access to information about out-of-sight friendly units. 
The observed vector by the agent includes relevant information about friendly and enemy units, such as distance, relative position, health, shield, and unit type. 
Additionally,  the regeneration of the shield after a period of time without attack and the requirement for armor to be broken before health decreases. 
The local observation of each agent is limited to its field of view, which encompasses the circular area of the map surrounding the unit and has a radius equal to the sight range, which is set to 6.
And the enemy army, controlled by a hand-coded built-in AI, presents a significant challenge with a difficulty level of 7. 
A detailed description of SMAC maps and experimental settings can be found in Table~\ref{table:kysymys}.

\begin{table}[hb]
	\centering
	\begin{tabular}{cccccc}
		\toprule
		Map            & Ally Units                                                                              & Enemy Units                                                                             & Difficulty & Steps & Exploration Steps \\
		\midrule
		8m             & 8 Marines                                                                               & 8 Marines                                                                               & Easy       & 2M    & 5000              \\
		2s3z           & \begin{tabular}[c]{@{}c@{}}2 Stalkers,\\      3 Zealots\end{tabular}                    & \begin{tabular}[c]{@{}c@{}}2 Stalkers,\\      3 Zealots\end{tabular}                    & Easy       & 2M    & 5000              \\
		3s5z           & \begin{tabular}[c]{@{}c@{}}3 Stalkers,\\      5 Zealots\end{tabular}                    & \begin{tabular}[c]{@{}c@{}}3 Stalkers,\\      5 Zealots\end{tabular}                    & Easy       & 2M    & 5000              \\
		2s\_vs\_1sc    & 2 Stalkers                                                                              & 1 Spine Crawler                                                                         & Easy       & 2M    & 5000              \\
		1c3s5z         & \begin{tabular}[c]{@{}c@{}}1 Colossi,\\      3 Stalkers,\\      5 Zealots\end{tabular}  & \begin{tabular}[c]{@{}c@{}}1 Colossi,\\      3 Stalkers,\\      5 Zealots\end{tabular}  & Easy       & 2M    & 5000              \\
\hline
		5m\_vs\_6m     & 5 Marines                                                                               & 6 Marines                                                                               & Hard       & 2M    & 5000              \\
		8m\_vs\_9m     & 8 Marines                                                                               & 9 Marines                                                                               & Hard       & 2M    & 5000              \\
		3s\_vs\_5z     & 3 Stalkers,                                                                             & 5 Zealots                                                                               & Hard       & 2M    & 5000              \\
		2c\_vs\_64zg   & 2 Colossi                                                                               & 64 Zerglings                                                                            & Hard       & 2M    & 5000              \\
\hline
		MMM2           & \begin{tabular}[c]{@{}c@{}}1 Medivac,\\      2 Marauders,\\      7 Marines\end{tabular} & \begin{tabular}[c]{@{}c@{}}1 Medivac,\\      3 Marauders,\\      8 Marines\end{tabular} & Super Hard & 2M    & 5000              \\
		6h\_vs\_8z     & 6 Hydralisks                                                                            & 8 Zealots                                                                               & Super Hard & 5M    & 10000             \\
		3s5z\_vs\_3s6z & \begin{tabular}[c]{@{}c@{}}3 Stalkers,\\      5 Zealots\end{tabular}                    & \begin{tabular}[c]{@{}c@{}}3 Stalkers,\\      6 Zealots\end{tabular}                    & Super Hard & 5M    & 10000             \\
		corridor       & 5 Zealots                                                                               & 24 Zerglings                                                                            & Super Hard & 5M    & 10000    \\ 
		\bottomrule
	\end{tabular}
	\caption{The introduction of maps and experimental settings in the SMAC benchmark.}
	\label{table:kysymys}
\end{table}

\section{Hyperparameter Settings}
\label{appendix2}
In our experiments, we adopt the Python MARL framework (PyMARL)~\cite{smac} to implement baselines including three representative value decomposition baselines, VDN~\cite{vdn}, QMIX~\cite{qmix} and QTRAN~\cite{qtran}, and four state-of-the-art MARL algorithms, including WQMIX~\cite{rashid2020wqmix}~(both CW-QMIX and OW-QMIX), QPLEX~\cite{qplex} CDS~\cite{li2021cds} and SHAQ~\cite{wang2021shaq}.
The hyper-parameters of the training and testing configurations are set in accordance with the configurations used in the SMAC framework and as described in the source codes, and are kept consistent across all baselines for fairness.
Our method is trained every 20,000 timesteps and evaluated using 32 episodes with decentralized greedy action selection to measure the test win rate of each algorithm. 
We train with 5 random seeds for environments with stark success or failure conditions.
This win rate is calculated as the percentage of episodes where our agents successfully defeated all enemy units within the time limit. 
In practice, we utilize the RMSprop optimizer with a learning rate of 0.0005 and update the target network parameters after every 200 episodes. 
Some of the commonly used experimental settings in multi-agent Q-learning are listed in Table~\ref{paramater1}.

Our UNSR is also based on the PyMARL implementation.
We manually encode the agent's observations as input and embed them into a 32-dimensional hidden space.
The individual network for each agent $i$ contains two layers of unit-wise state representation and three attention heads.
For the final embedding layer in the individual network, we also set 32 dimensions.
For the hybrid network, we set the global state $s$ and unit-wise state $\mathcal{Z}$ to a 32-dimensional hidden space and use four attention heads to obtain multiple generative vectors in parallel.
We summarise all the hyperparameters of the UNSR in Table~\ref{paramater2}.
The training was conducted on a single machine with an NVIDIA A5000 GPU and an Intel I9-12900k CPU. 
The total training time for our model varied depending on the difficulty of the experimental environment, with a range of 2 million to 5 million timesteps and a duration of 5 to 30 hours.

\begin{minipage}[c]{0.5\columnwidth}
\centering
\begin{tabular}{crllc}
\toprule
Hyper-parameter         & \multicolumn{1}{c}{} &  &  & Value   \\
\hline
Batch    Size                &                      &  &  & 32      \\
Test Interval           &                      &  &  & 2000    \\
Test Nepisode           &                      &  &  & 32      \\
TD Discount Factor                   &                      &  &  & 0.99    \\
Buffer Size             &                      &  &  & 5000    \\
Start Exploration Rate           &                      &  &  & 1       \\
End Exploration Rate             &                      &  &  & 0.05    \\
Target Update Interval  &                      &  &  & 200     \\
Grad Norm               &                      &  &  & 10      \\
Optimizer                &                      &  &  & RMSprop    \\
Learning Rate                       &                      &  &  & 0.0005  \\
\hline
RNN Hidden Dimension         &                      &  &  & 64      \\
Mixing Embed Dimension       &                      &  &  & 32      \\
Hypernet Embed Dimension        &                      &  &  & 64      \\
Number of Hypernet Layers         &                      &  &  & 2       \\
\bottomrule
\end{tabular}
\captionof{table}{The configurations of common settings.}
\label{paramater1}
\end{minipage}
\begin{minipage}[c]{0.5\columnwidth}
\centering
\begin{tabular}{ccccc}
\toprule
\begin{tabular}[c]{@{}c@{}}Individual   Network \\      Nyper-parameters\end{tabular} &  &  &  & Value \\
\hline
Unit Dimension                                                                       &  &  &  & 32    \\
Channel Dimension                                                                    &  &  &  & 32    \\
Unit-wise Space Dimension                                                           &  &  &  & 32    \\
Number of Attention Heads                                                                                 &  &  &  & 3     \\
Number of Representation Layers                                                                                 &  &  &  & 2     \\
\hline
\begin{tabular}[c]{@{}c@{}}Mixing  Network \\   Hyper-parameters\end{tabular}     &  &  &  & Value \\
\hline
State Embedding Dimension                                                      &  &  &  & 32    \\
Unit-wise Space Dimension                                                          &  &  &  & 32    \\
Number of Attention Heads                                                                              &  &  &  & 4     \\
Number of Representation Layers                                                                               &  &  &  & 1     \\
Number of Pooling Weight Layers                                                               &  &  &  & 2     \\
First Layer Weight Dimension   &  &  &  & 64    \\
Second Layer Weight Dimension                                                             &  &  &  & 4     \\
\bottomrule
\end{tabular}
\captionof{table}{The special hyper-parameters of UNSR architecture.}
\label{paramater2}
\end{minipage}

\section{The Full Comparison of UNST with SOTA on SMAC}
\label{appendix3}
We compare our UNSR with mainstream SOTA over almost hard and super hard scenarios.
The learning curves are shown as in Figure~\ref{all_results}.
The detailed parameter settings are shown as in Appendix~\ref{appendix2}.
Our UNSR surpasses existing SOTA by a large margin and achieves 100\% test win rates in almost all scenarios, especially on the super hard scenarios.
The superior representational capacity of UNSR combined with the unit-wise state representation and incorporating it into estimating joint value function presents a clear beneﬁt over existing value decomposition.
Therefore, UNSR achieves the new SOTA on the SMAC benchmark.

\begin{figure}[H]
  \centering
   \includegraphics[width=\columnwidth]{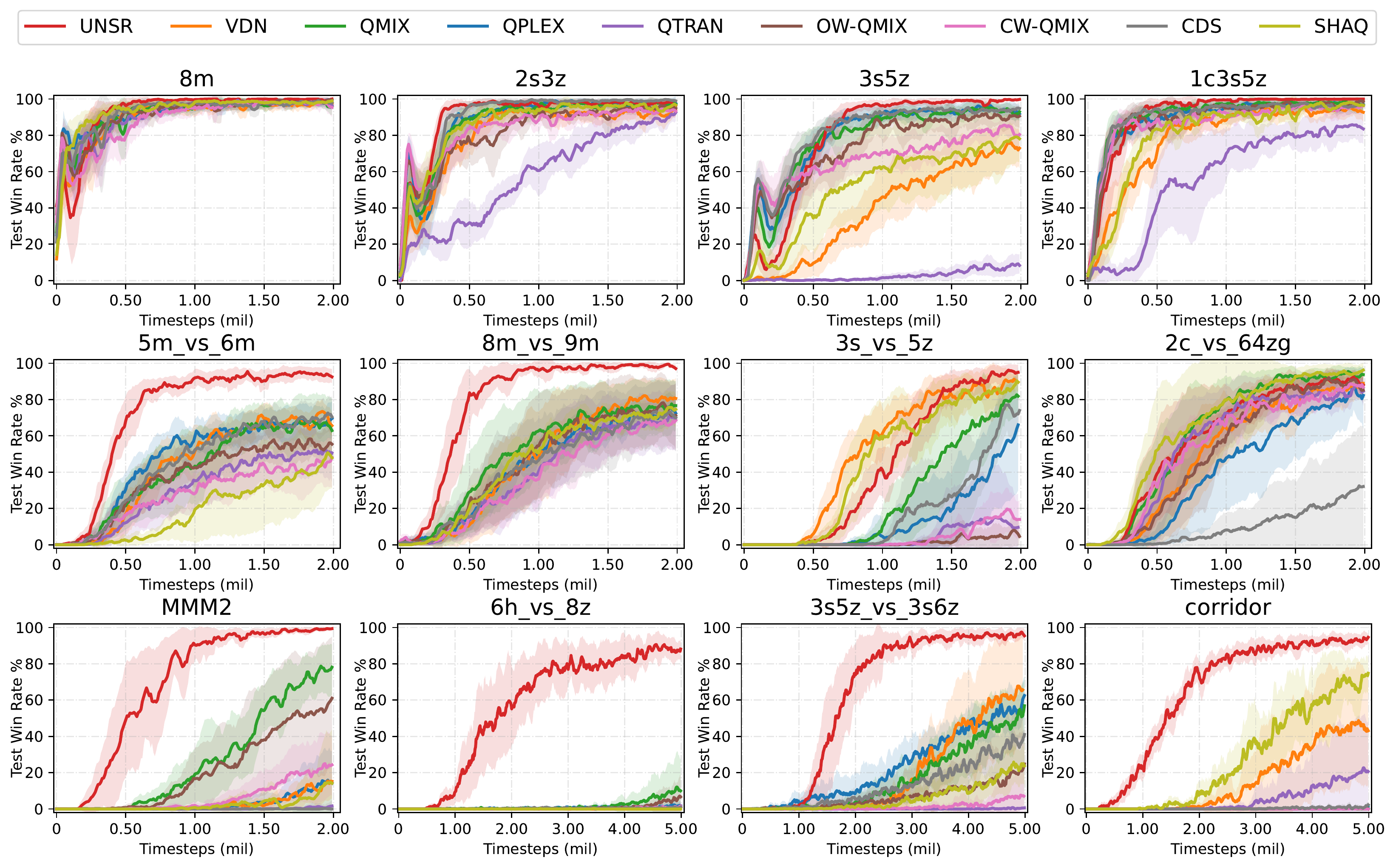}
   \caption{The learning curves of UNSR compared with the SOTA over hard and super hard scenarios.}
   \label{all_results}
\end{figure}

\end{document}